\documentclass[11pt]{article}
\usepackage{jheppub}

\usepackage{amsmath,graphicx,amssymb,bbm,subfigure,hyperref,array,slashed,cancel,physics,bbm}
\usepackage[all]{xy}
\usepackage{color}

\def\be{\begin{equation}}
\def\ee{\end{equation}}
\def\ba{\begin{eqnarray}}
\def\ea{\end{eqnarray}}

\def\p{\pi}

\def\IR{\relax{\rm I\kern-.18em R}}

\def\inv{^{\raise.0ex\hbox{${\scriptscriptstyle -}$}\kern-.05em 1}}

\title{A Type IIB Matrix Model via non-Abelian T-dualities} 

\author{Jeroen van Gorsel$^{1}$ and Salomon Zacar\' ias$^{1}$}

\affiliation{$^1$ Department of Physics, Swansea University, Swansea SA2 8PP, United Kingdom}

\emailAdd{jeroen.van.gorsel@gmail.com}
\emailAdd{salomon.zacarias@swansea.ac.uk} 

\abstract{ We construct a new solution of type IIB supergravity with eight supercharges by applying non-Abelian T-duality to the $ AdS_5\times S^5$ solution, along the $ SU(2)$ isometries in both the
 internal and Lorentz symmetries. The study of quantised charges and D-brane embeddings suggests a configuration of D1 and D3'-branes that polarise into 
 concentric, spherical, D3 and D5-branes due to the Myers dielectric effect. We find that our solution is dual to a IIB matrix model with fuzzy sphere vacua characterised by partitions of the number of 
 D1 and D3'-branes. We also study the solution obtained via (Abelian)
  T-dualities along the $S^1$ fiber angles inside the $S^3$'s  of the $ AdS_5\times S^5$ solution. In addition, we point out a precise relation between NATD and ATD solutions by considering a double scaling limit in the NATD solutions.
  In the particular case of the solutions obtained by a single (non-)Abelian T-duality inside the Lorentz symmetries of the $AdS_5\times S^5$ solution, we demonstrate that the double scaling limit provides a supergravity realisation of the relations 
  between the vacua of the half-BPS theories studied by Lin and Maldacena. 
\\[10pt]
 } 

\begin{document}
\def\Tr{{\textrm{Tr}}}

\maketitle



\section{Introduction}
In recent years, the notion of duality symmetries has prompted  a number of major, far reaching developments in theoretical physics. This concept provides a surprising link between physical systems which are apparently unrelated. 
In these cases, a system described in terms of a given set of variables, can also be understood in a simpler way when looked at in terms of a new, perhaps totally different, set. The story in the context of conventional field theories is very old and well-known \cite{AlvarezGaume:1997ix}. 
In string theory,  duality symmetries relate all of the different string theories  \cite{Polchinski:2014mva}. Moreover, in a seminal work \cite{tHooft:1973alw} 't Hooft  pointed out a relationship between gauge fields and strings. A particular realisation of this idea, found by Maldacena, relates the large N limit of superconformal field theories with strings moving in a curved background \cite{Maldacena:1997re}.
 The original example establishes a relation between strings propagating in the background $AdS_5\times S^5$ and the $\mathcal{N}=4$ SYM theory living at the boundary of this spacetime. Since then, an impressive amount of work has appeared in the literature, aimed at finding new examples of this gauge/gravity correspondence \cite{Aharony:1999ti}. 
Diverse and sophisticated techniques have been developed to construct and classify new solutions in supergravity with the identification of their field theory duals being of paramount importance.
In this vein, symmetries play an important role as they strongly constrain the form of solutions whilst their stability is guaranteed by preserving some fraction of supersymmetry.

Moreover, duality symmetries have also been used to produce new string backgrounds. Notably, non-Abelian T-duality (NATD), which is
the generalisation of the well-known Abelian T-duality (ATD) to non-Abelian isometries. In concrete, T-duality states the equivalence of strings propagating in spacetimes with small and large compact directions (circles). In a sense, NATD is the generalisation of this idea to more complicated compact spaces.
The NATD story started with \cite{delaOssa:1992vci}, where the T-duality transformations for the NS-sector of string theory were generalised to the case of non-Abelian isometries. Later, in the work of \cite{Sfetsos:2010uq}, these transformations were extended by incorporating the transformation of the RR fields, providing a method to generate new solutions in type II supergravity for backgrounds with non-Abelian isometries. Since then, a substantial amount of work constructing new solutions of type II supergravity generated via NATD - which are difficult to find by conventional methods - has appeared in the literature \cite{variosa1, variosa4, variosa5, variosb1}.
It is worth pointing out that presently the target space duality character of NATD has not been formally proven. Instead,  it is seen as an operation relating different theories, as has been noticed by studying the field theory duals of certain supergravity solutions obtained via NATD.

This development started in the work of Lozano and N{\'u}{\~n}ez \cite{Lozano:2016kum} where the role of NATD in a field theoretical context was investigated. The example studied in  \cite{Lozano:2016kum}  was the solution generated via NATD along the $SU(2)$ isometries inside $S^5$ of the $AdS_5\times S^5$ background, the Sfetsos-Thompson solution \cite{Sfetsos:2010uq}. There it was found that the dual field theory to this configuration corresponds to an $\mathcal{N}=2$ superconformal field theory, described by a linear quiver of infinite length. Interestingly, it was noticed that the matching of observables between field theory and the supergravity configuration
 imposes a completion of the solution in such a way that certain global issues inherent to the NATD geometry, like singularities and the boundedness of the T-dual coordinates, can be addressed.
The results of \cite{Lozano:2016kum} were further elaborated to propose field theory duals to other solutions obtained via NATD in \cite{Lozano:2016wrs,Lozano:2017ole, Itsios:2017cew}.
A common feature of these works is that the NATD solutions can be understood as a patch of well-defined corresponding geometries.
 An interesting realisation of this idea was found in \cite{Lozano:2017ole} where it was shown that the solution obtained by applying NATD to the $AdS_5\times S^5$ solution along the SU(2) isometries inside $AdS_5$  corresponds, when uplifted to eleven dimensions, to the Penrose limit of the $AdS_7 \times S^4$ superstar solution.

In this work we will exploit the application of two (N)ATD's to generate new solutions of type IIB supergravity with 8 supercharges. 
The starting point will be to consider the  NATD solution of $AdS_5\times S^5$ obtained after a single NATD applied along the $SU(2)$ isometries inside $AdS_5$, the field theory dual of which 
corresponds to a particular vacuum of the BMN matrix model \cite{Lozano:2017ole}. 
Following the ideas developed in \cite{Lozano:2016kum}, we will demonstrate that the dual field theory of the new solution corresponds to a matrix model in type IIB supergravity, characterised by fuzzy sphere vacua
which, on the supergravity side, correspond to D1 and D3'-branes that polarise into D3 and D5-branes respectively.
We will see that this field theory shares some similarities with the BMN matrix model.
This is natural since we naively expect that the role played by D0-D2 branes  in the BMN matrix model will map to a system of
D3'-D5 and D1-D3 branes after applying the NATD.
We will support these ideas by computing D-brane embeddings that show that the D1 and D3'-branes blow up into D3 and D5-branes with a radius proportional to their charge.

Moreover, by T-dualising the $AdS_5\times S^5$ solution along the Hopf-fibre angle on the $S^3$ inside $AdS_5$, after considering large gauge transformations,
  we will see that the solution fits into the Lin-Maldacena half-BPS class of solutions in type IIA supergravity \cite{Lin:2004nb}. The dual field theory is therefore $\mathcal{N}=4$ SYM on $\mathbb{R}\times S^3/\mathbb{Z}_{k}$.   
 The application of a second ATD along the Hopf-fibre angle inside $S^5$ now generates a IIB solution characterised by two sets of NS5-branes periodically arranged along the Hopf-fibre dual angles with D3-branes stretched between
  one of these sets of NS5-branes. Further, we will point out a precise relation between NATD and ATD solutions by taking a double asymptotic limit in the NATD solutions. 
  In the particular case of the NATD and ATD solutions obtained by applying 
  a NATD along the SU(2) and Hopf-fibre T-duality inside $AdS_5$ we take this double asymptotic limit in the potential function -in the Lin-Maldacena language- characterising these half-BPS solutions.
  The above relation between solutions is therefore   a supergravity realisation of the relations between the field theory duals to the Lin-Maldacena class of geometries \cite{Lin:2005nh,Ishiki:2006yr}.

 The organisation of this paper is as follows.
 \begin{itemize}
 \item In section \ref{sec:AdssxS5andNATDSolutions} we briefly review the solutions obtained via a single NATD on the $AdS_5\times S^5$ solution along the $SU(2)$ isometries inside either the $AdS_5$ or $S^5$ subspaces. 
 \item In section \ref{sec:DoubleNATDBackground} we present a new solution of type IIB supergravity, obtained by applying NATD to the solutions studied in section \ref{sec:AdssxS5andNATDSolutions}.
We notice that the application of NATD is commutative. In section \ref{subsec:QuantisedCharges} we study the properties of our solution, studying the quantised charges taking into account large gauge transformations. We will see that the brane set-up characterising the solution is given by an intersection of a
D1-D3-D3'-D5-NS-NS' system, the low energy description of which is dual to a IIB matrix model characterised by fuzzy sphere vacua. In the supergravity side this is realised by D1 and D3'-branes blowing up into D3 and D5-branes due to the Myers dielectric effect. 
\item In Section \ref{sec:DielectricBranes} we study the D-brane embeddings corresponding to Myers dielectric D3 and D5-branes. 
\item  In section \ref{subsec:ATDonAdS5} we study the ATD solution obtained by applying an ATD along the Hopf fibre angle  inside $AdS_5$ of the $AdS_5\times S^5$ solution and see that it fits into the Lin-Maldacena class of geometries. The application of a subsequent ATD  along the Hopf fibre angle  inside $S^5$  is considered in
\ref{subsec:DATD}. It gives rise to a new solution of type IIB supergravity characterised by periodic arrangements of  NS$5_1$ and NS$5_2$ fivebranes along the dualised Hopf-fibre angles with 
  D3-branes stretched between the NS$5_1$ fivebranes. We also discuss a double scaling limit relating NATD and ATD solutions.
  
  \item In section \ref{sec:Conclusion} we present some concluding remarks.
\item We have also included two Appendices. In Appendix \ref{sec:susy} we show that an spinor with 8 supercharges is consistent with the independence  
  on the SU(2) Euler angles of the $AdS_5\times S^5$ spinor and according with the results in \cite{Kelekci:2014ima} and \cite{Hassan:1999bv}, our solutions preserves eight supercharges. 
  In Appendix \ref{scaling} we study a relation via a double asymptotic limit between the NATD and ATD solutions studied in sections \ref{LNsolution} and \ref{subsec:ATDonAdS5}
   in the context of the Lin-Maldacena class of geometries.
   \end{itemize}

\section{$\text{AdS}_5 \times \text{S}^5$ and NATD Solutions}\label{sec:AdssxS5andNATDSolutions}
In this section we start by giving a brief overview of the supergravity solutions that we will refer back to throughout the rest of this paper. We refer the reader to \cite{Lozano:2016kum, Lozano:2017ole} for a more detailed explanation of their properties. All of the solutions that we will discuss here are obtained by applying NATD's (or ATD's in section \ref{DATD}) to the well-known type IIB $AdS_5\times S^5$ solution. The metric of this solution is 
\begin{equation}\label{adsspace}
   ds^2 = L^2\left(ds^2_{AdS_5}+d\Omega_5^2\right),
\end{equation}
where, in global coordinates, the line elements are
\begin{align}
     ds^2_{AdS_5}&=-\cosh^2 r dt^2 + dr^2 + \frac{\sinh^2 r}{4}\left(\omega_1^2 + \omega_2^2 + \omega_3^2 \right), \\
    d\Omega_5^2 &= d\alpha^2 + \sin^2 \alpha d\beta^2 + \frac{\cos^2 \alpha}{4}(\sigma_1^2 + \sigma_2^2 + \sigma_3^2).
\end{align}
Here $\omega_{i}$ and $\sigma{_i}$ are two sets of left invariant Maurer-Cartan forms parametrising the $SU(2)$ isometries inside the $AdS_5$ and $S^5$ subspaces respectively. They are explicitly given by 
\begin{equation}\label{mcforms}
\begin{split}
\omega_1=&\cos\psi\sin\theta_1 d\phi_1-\sin\psi_1 d\theta_1,\qquad \sigma_1=\cos\psi\sin\theta d\phi-\sin\psi d\theta,\\
\omega_2=&\sin\psi_1\sin\theta_1 d\phi_1-\cos\psi_1 d\theta_1,\quad~~ \sigma_2=\sin\psi\sin\theta d\phi-\cos\psi d\theta,\\
\omega_3=&d\psi_1+\cos\theta_1 d\phi_1,\qquad \qquad  \qquad \quad \sigma_3=d\psi+\cos\theta d\phi,
\end{split}
\end{equation}
where $\alpha \in [0, \frac{\pi}{2}]$, $\beta \in [0, 2\pi]$, $\phi, \phi_1 \in [0, 2\pi]$, $\theta$, $\theta_1\in [0, \pi]$ and $\psi,\psi_1\in [0, 4\pi]$. This solution is supported by a self-dual RR five-form field strength,
\begin{equation}\label{flux}
F_5=\frac{4}{L}(\textrm{Vol}(AdS_5)-\textrm{Vol}(\Omega_5)),
\end{equation}
with $N_{D_3}$ units of flux\footnote{Throughout this paper we will be using $2\kappa_{10}^2 T_{D_p}=(2\pi)^{7-p}\alpha'^{(7-p)/2}$ and $g_s=1$.} 
\begin{equation}
N_{D_3}=\frac{1}{2\kappa_{10}^2 T_{D_3}}\int_{S^5} F_{5}, \qquad L^{4}=\frac{\pi}{4}g_s^2 N_{D_3}\alpha'^2.
\end{equation}
The bosonic symmetries of this solution are $SO(4,2)\times SO(6)$ and preserves 32 supercharges.

\subsection{NATD inside $S^5$}\label{ST}\label{sfetsos-thompson}
In \cite{Sfetsos:2010uq} a NATD was applied along the $SU(2)\subset SO(6)$ isometries that are parametrised by the $\sigma_{i}$'s in eq. (\ref{mcforms}) of the $AdS_5 \times S^5$ solution in eqs. (\ref{adsspace}) and (\ref{flux}). The resulting type IIA solution reads \footnote{Notice that we are using a different normalisation to the one originally presented in \cite{Sfetsos:2010uq}.}
\begin{equation}\label{natd1}
\begin{split}
   ds^2 &=L^2 ds_{AdS_5}^2+L^2(d\alpha^2+\sin^2\alpha\;d\beta^2)+\frac{4\alpha'^2}{L^2 \cos^2\alpha}d\rho^2+\frac{4 L^2 \alpha'^2\rho^2\cos^2\alpha}{\Delta}d\Omega_2^2(\chi,\xi)\\
    &\quad\qquad B_2 =- \frac{16 \alpha'^3\rho^3}{\Delta} d\Omega_2(\chi,\xi),\quad 
   e^{-2\phi} = \frac{L^2\cos^2\alpha}{64\alpha'^3}\Delta, \\
   &\quad\quad\quad F_2= \frac{L^4}{2\alpha'^{3/2}}\cos^3\alpha\sin\alpha\;d\alpha\wedge d\beta, \quad F_4=B_2\wedge F_2,
   \end{split}
\end{equation}
where 
\begin{equation}\label{delta}
\Delta=16\alpha'^2\rho^2+L^4\cos^4\alpha.
\end{equation}
This solution is singular at $\alpha=\pi/2$ due to the presence of NS5-branes and belongs to the Gaiotto-Maldacena class of geometries \cite{Sfetsos:2010uq, Lozano:2016kum}. 

  \subsection{NATD inside $AdS_5$}\label{LNsolution}
When we apply a NATD on the non-compact subspace of the $AdS_5 \times S^5$ solution (\ref{adsspace}) along the $SU(2)$ parametrised by the $\omega_i$'s in eq. (\ref{mcforms}) we find a solution of type IIA given by \cite{Lozano:2017ole}
 \begin{equation}\label{natd2}
 \begin{split}
   ds^2 &=  L^2( -\cosh^2 rdt^2 + dr^2) + \frac{4 {\alpha'}^2}{L^2 \text{sinh}^2r} d\rho_1^2 + \frac{4 L^2 {\alpha'}^2 \rho_1^2 \sinh ^2 r}{\tilde{\Delta}} d\Omega_2^2(\chi_1, \xi_1) +L^2 d\Omega_5^2,\\
     &\quad\qquad B_2 = -\frac{16 {\alpha'}^3\rho_1^3 }{\tilde{\Delta}} d\Omega_2(\chi_1, \xi_1) ,\quad e^{-2\phi} = \frac{L^2 \sinh^2 r}{64 \alpha'^3} \tilde{\Delta},\\
   &\quad\quad\quad F_2= \frac{L^4}{2\alpha'^{3/2}}\cosh r\sinh^3 r\;dr\wedge dt,\quad F_4 = B_2 \wedge F_2,
   \end{split}
\end{equation}
where
\begin{equation}\label{deltat}
\tilde{\Delta}=16 {\alpha'}^2 \rho_1^2 + L^4 \sinh^4 r.
\end{equation}
The singularity of this solution at $r=0$ is, as in the case of the previous solution (\ref{natd1}), due to the presence of NS5-branes. This solution has $\mathbb{R}_t \times SO(3) \times SO(6)$ isometries and fits into the Lin-Maldacena classification of half-BPS solutions of type IIA supergravity \cite{Lin:2005nh}. This can be easily seen from the fact that via an analytical continuation \cite{Lozano:2017ole} the solution in eq. (\ref{natd1}) maps to the one in eq. (\ref{natd2}). However, it is worth noticing that this relation should be taken carefully since both solutions are rather different. Namely, whilst the solution in eq. (\ref{natd2}) is dual to a particular vacuum of the BMN matrix model \cite{Lozano:2017ole} the one in eq. (\ref{natd1}) is dual to a 4D SCFT with a quiver structure \cite{Lozano:2016kum}.

\section{The Double NATD of $AdS_5$ and $S^5$}\label{sec:DoubleNATDBackground}
Given the description of the above geometries that are obtained after applying one NATD in terms of half-BPS geometries that fit into different classifications, it is natural to consider the application of a second NATD in the residual $SU(2)$ isometries of these solutions. Let us start with the solution in equation (\ref{natd2}) and perform a NATD along the $SU(2)$ isometries inside the $S^5$ parameterised by the Mauer-Cartan forms $\sigma_i$ defined in eq. (\ref{mcforms}). The resulting background is a solution of type IIB supergravity and has an NS sector given by 
\begin{equation}\label{newNATD}
\begin{split}
   ds^2 &=  L^2\left( -\cosh^2 rdt^2+dr^2 + \frac{4 {\alpha'}^2}{L^4 \text{sinh}^2r} d\rho_1^2 + \frac{4 {\alpha'}^2 \rho_1^2 \sinh ^2 r}{\tilde{\Delta}} d\Omega_2^2(\chi_1,\xi_1)\right.\\
   &\quad\qquad\left. +d\alpha^2+\sin^2 \alpha d\beta^2 + \frac{4 \alpha'^2}{L^4 \cos^2 \alpha}d \rho^2 + \frac{4 {\alpha'}^2 \rho^2 \cos^2 \alpha}{\Delta}d\Omega_2^2(\chi,\xi)\right),\\
   B_2 &=- \frac{16 {\alpha'}^3 \rho^3}{\Delta}d\Omega_2(\chi,\xi) - \frac{16 {\alpha'}^3\rho_1^3}{\tilde{\Delta}}d\Omega_2(\chi_1,\xi_1),\qquad 
   e^{-2\phi} = \frac{L^4 \cos^2 \alpha \sinh^2 r}{(64 \alpha'^3)^2}\Delta \tilde{\Delta},
   \end{split}
\end{equation}
where the functions $\Delta$ and $\tilde{\Delta}$ were defined in eqs. (\ref{delta}) and (\ref{deltat}) respectively.
This solution has two singular points at $r=0$ and $\alpha=\frac{\pi}{2}$ that will correspond to two sets of NS and NS' fivebranes equally separated along the $\rho$ and $\rho_1$ axes, respectively. This fact will be verified later by studying the quantised charges of the solution. The RR sector supporting this solution is 
\begin{equation}
  \begin{split}\label{newNATDrr}
F_3 &= - \frac{L^4}{2\alpha'} \left(\rho \cosh r\sinh^3 r\;dr\wedge dt\wedge d\rho +\rho_1 \sin \alpha \cos^3 \alpha \;d\alpha\wedge d\rho_1\wedge d\beta\right),\\
   F_5 &= \frac{L^8 \rho^2 \cos ^4 \alpha \sinh ^3 r \cosh r}{2\Delta} dt\wedge dr\wedge d\rho \wedge d\Omega_2(\chi, \xi)\;+\\
  & \quad\frac{8 \alpha'^2 L^4 \rho  \rho_1^3 \sinh ^3 r \cosh r}{\tilde{\Delta}}\;dr\wedge dt\wedge d\rho\wedge d\Omega_2(\chi_1, \xi_1)\;+\\
   &\quad  \frac{8\alpha'^2 L^4 \rho ^3 \rho_1 \sin \alpha \cos ^3 \alpha}{\Delta}\;d\alpha\wedge d\rho_1\wedge d\beta\wedge d\Omega_2(\chi, \xi)\;-\\
   &\quad \frac{ L^8 \rho_1^2 \sin \alpha \cos ^3 \alpha \sinh ^4 r}{2\tilde{\Delta}}d\alpha\wedge d\rho_1\wedge d\beta\wedge d\Omega_2(\chi_1,\xi_1),\\
   F_7&= - \frac{8 L^8 \alpha'^3 \rho^2\rho_1^3 \cos^4\alpha \cosh r\sinh^3 r}{\Delta \tilde{\Delta}} dt\wedge dr\wedge d\rho \wedge d\Omega_2(\chi, \xi) \wedge d\Omega_2(\chi_1, \xi_1)\;+ \\
   &\quad \frac{8 L^8 \alpha'^3 \rho^3\rho_1^2 \cos^3\alpha\sin\alpha \sinh^4 r}{\Delta \tilde{\Delta}} d\beta\wedge d\alpha\wedge d\rho_1 \wedge d\Omega_2(\chi, \xi) \wedge d\Omega_2(\chi_1, \xi_1)
\end{split}
\end{equation}
Notice that the RR-sector of the solution is rather non-trivial as it mixes both dualised spaces. The solution is $\mathbb{R}_t\times SO(3)\times SU(2)\times U(1)$ symmetric  and preserves 8 supercharges  (see Appendix \ref{sec:susy}). 
Had we started with the solution in eq. (\ref{natd1})
 and applied a NATD along the $SU(2)$ isometries inside $AdS_5$ we would have obtained the same solution in eqs. (\ref{newNATD})-(\ref{newNATDrr}) as a consequence that the application of NATD is commutative.

For future reference we write the RR potentials associated with this solution satisfying $F_{p+1}=dC_p-H_{3}\wedge C_{p-2}$. They are 
\begin{align}
C_2 =& \frac{L^4}{8 \alpha'}\left(-\rho \sinh^4 r dt \wedge d \rho + \rho_1 \cos^4 \alpha d\rho_1 \wedge d\beta \right)\label{c2},\\
C_4 =& \frac{L^4 }{8}\alpha'^2 \rho\;\rho_1\left(\frac{\rho_1^2}{\tilde{\Delta}} \sinh^4 r\;dt\wedge d\rho \wedge d\Omega_2(\chi_1, \xi_1) + \frac{\rho^2}{\Delta} \cos^4 \alpha\;d\beta\wedge d\rho_1 \wedge d\Omega_2(\chi, \xi)\right)\nonumber\\
-& \frac{L^8 }{8}\cos^4 \alpha\;\sinh^4 r\;\left(\frac{\rho^2}{\Delta}\;dt\wedge d\rho \wedge d\Omega_2(\chi, \xi) + \frac{\rho_1^2}{\tilde{\Delta}}\;d\beta\wedge d \rho_1 \wedge d\Omega_2(\chi_1, \xi_1)\right), \label{c4}\\
C_6 =& 2L^6 \alpha'^2 \rho \rho_1 \Big(\frac{\rho^2_1}{\tilde{\Delta}} \sinh^3 r\; \cosh r\; \cot \alpha\;dt\wedge dr\wedge d\rho\wedge d\alpha\wedge d\Omega_2(\chi_1, \xi_1)\nonumber\\ 
& \qquad\qquad + \frac{\rho^2}{\Delta} \cos^3 \alpha\; \sin \alpha\; \tanh r\;d\beta \wedge d \alpha\wedge d\rho_1 \wedge dr \wedge d\Omega_2(\chi, \xi) \Big) \label{c6}\\
&+ \frac{2 L^8 \alpha'^3}{\Delta \tilde{\Delta}}\rho^2 \rho_1^2\;\cos^4\alpha\;\sinh^4 r \Big( \rho\;d\beta\wedge d\rho_1 \wedge d\Omega_2(\chi, \xi)\wedge d\Omega_2(\chi_1, \xi_1) \nonumber\\
&\qquad\qquad\qquad\qquad\qquad\qquad +\rho_1\;dt\wedge d\rho \wedge d\Omega_2(\chi, \xi)\wedge d\Omega_2(\chi_1, \xi_1)\Big). \nonumber
\end{align}
Let us now proceed to the study of the Page charges which will allow us to identify the properties of our solution in detail.

\subsection{Quantised Charges}\label{subsec:QuantisedCharges}
As in many other examples which have worked out NATD solutions and their field theory realisations, in this section we shall study the information obtained from the Page charges
of the solution in eqs. (\ref{newNATD}) and (\ref{newNATDrr}).

We start by noticing that a common feature of many $SU(2)$ NATD solutions is the presence of singular points in the geometry.
Such singularities correspond to the points where the $S^3$ along we are dualising the background shrinks to zero size and are
 originated from the presence of NS5-branes. For the solution in eq. (\ref{newNATD}) we see that there are two such singular points, at $r=0$ and at $\alpha=\frac{\pi}{2}$.  The leading order behaviour of the metric and dilaton close to these points are 
\begin{align}\label{intersecting}
ds^2=&L^2\left(-dt^2+d\beta^2+\frac{1}{4\nu}\left(d\tilde{\rho_1}^2+d\nu^2+\nu^2 d\Omega_2^2(\chi_1,\xi_1)\right)+\frac{1}{4\mu}\left(d\tilde{\rho}^2+d\mu^2+\mu^2 d\Omega_2^2(\chi,\xi)\right)\right),\nonumber\\
&\qquad \qquad \qquad \qquad \qquad \qquad \qquad e^{\Phi}\sim \frac{4\alpha'}{L^2\sqrt{\mu \nu}}\frac{1}{\rho\rho_1},
\end{align}
where $\nu=r^2,\mu=(\alpha-\frac{\pi}{2})^2$, $\tilde{\rho}=16\alpha'^2/L^4\rho$ and the same for $\tilde{\rho}_1$. The metric in eq. (\ref{intersecting}) can be thought of as a continuous distribution
of NS and NS' fivebranes along the $\tilde{\rho_1}$ and $\tilde{\rho}$ directions. This can also be verified by measuring the units of $H_3$ flux through the relevant cycles. Close to $r=0, \alpha=\pi/2$ the leading order behaviour of the $B_2$ field is $B_2\sim \alpha'\rho_1\;d\Omega_{2}(\chi_1,\xi_1)+\alpha'\rho\;d\Omega_2(\chi,\xi)$.
In \cite{Lozano:2013oma,Lozano:2014ata} an argument was proposed to bound the non-compact coordinate appearing in the NATD solutions, in the case at hand $\rho_1$ and $\rho$, such that the number of fivebranes is finite. The argument relies on the boundedness of the quantity 
\begin{equation}
b_0=\frac{1}{4\pi^2\alpha'}\int_{\Sigma}B_2 \quad  \in [0,1].
\end{equation}
In the present case we have two non-trivial two-cycles given by  $S^2_{(\chi_1,\xi_1)}$ and $S^2_{(\chi,\xi)}$ close to $r=0$ and $\alpha=\frac{\pi}{2}$ respectively.
Hence, according to \cite{Lozano:2013oma,Lozano:2014ata}, in order to achieve $b_{0}\in [0,1]$ over the above two cycles we have to impose simultaneously that $\rho_1\in [0,\pi]$ and $\rho\in [0,\pi]$, which makes the $(\rho_1,\rho)$-plane a grid of size $\pi$, the boundary of which is delimited by NS5-branes, denoted by NS5 and NS5'.  We then see that in order to fully cover the non-compact range of $\rho,\rho_1 \in \mathbb{R}^+$ and keep $b_0 \in [0,1]$ , a large gauge transformation of the form 
\begin{equation}
B_2 \to \hat{B}_2 = B_2-n_{1}\pi d\Omega_{2}(\chi_1,\xi_1)-n\pi d\Omega_2(\chi,\xi),
\end{equation}
is required whenever we pass through the intervals $[n_1\pi,(n_1+1)\pi]$ and $[n\pi,(n+1)\pi]$. 
Therefore, the $H_3$ flux through the cycles $(\rho_1, S^2_{(\chi_1,\xi_1)})\vert_{r=0}$ and $(\rho,  S^2_{(\chi,\xi)})\vert_{\alpha=\frac{\pi}{2}}$ is, respectively, 
\begin{equation}
    N_{_{NS5}} = \frac{1}{4\pi^2\alpha'}\int_{S^2_{(\chi_1, \xi_1)}}\int_{0}^{n_1\pi }\;H_{3}=n_1,\qquad
    N_{_{NS5'}} = \frac{1}{4\pi^2\alpha'}\int_{S^2_{(\chi, \xi)}}\int_{0}^{n\pi}\;H_3=n.
\end{equation}
We then have NS5-branes located at positions $\rho_1=\pi,2\pi\ldots n_1\pi $ with their worldvolume along $(t,\alpha,\beta,\rho,\chi,\xi)$ and  NS5' branes at $\rho=\pi,2\pi\ldots n\pi $ extended along $(t,\beta,r,\rho_1,\chi_1,\xi_1)$.
These two sets of NS5-branes divide the entire $(\rho_1,\rho)$-plane into a grid. Notice that, the above discussed configuration of NS5-branes in $(\rho_1,\rho)$ is a consequence that the $B_2$ field in
 (\ref{newNATD}) is a superposition of the $B_2$ fields in (\ref{natd1}) and (\ref{natd2}).\\

In addition we have non-zero $F_3$, $F_5$ and $F_7$ RR field strengths, so that we can measure the flux of these fields through non-trivial cycles  present in the geometry. We have four piecewise compact cycles defined by 
\begin{equation}
\Sigma_3=(\rho_1,S^2_{(\alpha,\beta)}), \quad \Sigma'_{5}=\Sigma_3\otimes S^2_{(\xi_1,\chi_1)}, \quad  \Sigma_{5}=\Sigma_3\otimes S^2_{(\xi,\chi)}, \quad \Sigma_7=\Sigma_3\otimes S^2_{(\xi_1,\chi_1)}\otimes S^2_{(\xi,\chi)}.
\end{equation}
The D5 Page charge computed in the interval $\rho_1 \in [n_1\pi, (n_1+1)\pi]$ is
\begin{align}
  N_{D5} =& \frac{1}{4 \pi^2\alpha'}\int_{\Sigma_3} F_3
    =\frac{L^4 \pi}{32 \alpha'^2}(1+2n_1),
\end{align}
where the D5-branes are extended along $(t,\rho,\chi_1,\xi_1,\chi, \xi)$. Moreover, Page charge for D1 and D3-branes
is induced due to large gauge transformations on the $B_2$ field. In the interval $\rho_1 \in [n_1\pi, (n_1+1)\pi]$ and $\rho \in [n\pi, (n+1)\pi]$, these Page charges are
\begin{align}
    N_{D3'} =& \frac{1}{(2\pi)^4\alpha'^{2}}\int_{\Sigma'_5} F_5 - F_3 \wedge \hat{B}_2
    = \frac{L^4 \pi}{48 \alpha'^2} \left(3n_1^2 + 3n_1 + 1\right) + n_1 N_{D5},\label{disolved}\\
    N_{D3}=& \frac{1}{(2\pi)^4\alpha'^{2}}\int_{\Sigma_5} F_5 - F_3 \wedge \hat{B}_2 = n \frac{L^4 \pi}{32 \alpha'^2}(1+2n_1)=n N_{D5},\label{comparison}\\
    N_{D1} =&\frac{1}{(2\pi)^6\alpha'^{3}}\int_{\Sigma_7}F_7-F_{5}\wedge \hat{B}_2+\frac{1}{2}F_3\wedge \hat{B}_2\wedge \hat{B}_2=n_1N_{D_3}+ n\frac{L^4\pi}{48 \alpha'^{2}}(1-3n_1^2)\label{d1}.
\end{align}
We find there are two sets of D3-branes, one with worldvolume coordinates $(t,\rho,\chi,\xi)$ (denoted by D3'), and one along $(t,\rho,\chi_1,\xi_1)$ (denoted by D3). From eqs. (\ref{comparison}) and (\ref{d1}) we see there is a contribution of the D1-branes to the D3 charge coming from the term,
\begin{equation}\label{disd3}
N_{D1}^{(d)}= \frac{n n_1 }{2^6\pi^4\alpha'^{3}}\int_{\Sigma_5} F_{3(\rho_1,\alpha,\beta)}d\Omega_2(\chi,\xi)\int_{S^{2}_{(\xi_1,\chi_1)}} d\Omega_2(\chi_1,\xi_1)=n_1 N_{D3}.
\end{equation}
Therefore we see that this fraction of the D1 charge is dissolved into the D3-branes. This suggest that we have D1-branes expanded onto spherical D3-branes due to the presence of the $B_{2(\chi_1,\xi_1)}$ field, as a consequence of the Myers dielectric effect \cite{Myers:1999ps}. The rest of the charge is usual D1-brane charge.
Notice that, from eq. (\ref{comparison}), one might think that the D3-branes will, in addition, blow up into D5-branes. This, however, is not possible as the D3 would then blow up onto an $\Omega_2(\chi, \xi)$ 2-sphere of radius $\rho$, which is not possible, as the $\rho$-coordinate is part of the worldvolume of the D3-branes. 
This effect is similar to the one found in \cite{Lozano:2017ole} in which D0-branes expanded onto fuzzy two-spheres to form spherical D2-branes. In analogy to this case, we will identify below $\rho_1$ as the radius of the expanded D1-branes into $S^2_{(\chi_1,\xi_1)}$.

Moreover, the number $N_{D3'}$ of D3' branes increases every time we go through the interval $\rho_1 \in [n_1\pi, (n_1+1)\pi]$. From eq. (\ref{disolved}) we see that a fraction,
\begin{equation}\label{disd5}
 N_{D3'}^{(d)}=\frac{1}{(2\pi)^4\alpha'^{2}}\int_{\Sigma'_5} F_{3(\rho_1,\alpha,\beta)}\int_{S^{2}_{(\xi_1,\chi_1)}} d\Omega_2(\chi_1,\xi_1)=n_1 N_{D5},
\end{equation}
of this D3' charge is dissolved in the D5-branes. Again, the remaining charge is usual D3' charge.
From eqs. (\ref{disd3}) and (\ref{disd5}),  the total, dissolved, charge carried by $k$ concentric D3 and D5-branes is 
\begin{equation}
N_{D1}^{(d)}=\sum_{n_1=1}^{k} n_1 N_{D_3},\quad N_{D3'}^{(d)}=\sum_{n_1=1}^{k} n_1 N_{D_5}.
\end{equation}
The above relations indicate that the vacua of the theory correspond to reducible $SU(2)$ representations where $n_1$, associated to the radius of the spherical branes, corresponds to the
 dimension of the representation whilst  $N_{D_3}$ and $N_{D_5}$ are their multiplicities. This is similar to the vacua of the $\mathcal{N}=1^{*}$ theory studied by Polchinsky and Strassler \cite{Polchinski:2000uf}.

 \begin{table}[h!]
\centering
 \begin{tabular}{c||c c|c c c|c c|c c c}
  & $t$ & $r$ & $\rho_1$ & $\chi_1$ & $\xi_1$ & $\alpha$ & $\beta$ & $\rho$ & $\chi$ & $\xi$ \\ [0.5ex] 
 \hline\hline 
 NS5' & $\bullet$ & $\bullet$ & $\bullet$ & $\bullet$ & $\bullet$ & $\cdot$ & $\bullet$ & $\cdot$ & $\cdot$ & $\cdot$ \\ 
 \hline
 NS5 & $\bullet$ & $\cdot$ & $\cdot$ & $\cdot$ & $\cdot$ & $\bullet$ & $\bullet$ & $\bullet$ & $\bullet$ & $\bullet$ \\ 
 \hline\hline 
  D5 & $\bullet$ & $\cdot$ & $\cdot$ & $\bullet$ & $\bullet$ & $\cdot$ & $\cdot$ & $\bullet$ & $\bullet$ & $\bullet$ \\
  \hline
  D3' & $\bullet$ & $\cdot$ & $\cdot$ & $\cdot$ & $\cdot$ & $\cdot$ & $\cdot$ & $\bullet$ & $\bullet$ & $\bullet$ \\
 \hline\hline 
  D3 & $\bullet$ & $\cdot$ & $\cdot$ & $\bullet$ & $\bullet$ & $\cdot$ & $\cdot$ & $\bullet$ & $\cdot$ & $\cdot$ \\ 
 \hline
  D1 & $\bullet$ & $\cdot$ & $\cdot$ & $\cdot$ & $\cdot$ & $\cdot$ & $\cdot$ & $\bullet$ & $\cdot$ & $\cdot$
\end{tabular}
\caption{Overview of the brane content of the Type IIB supergravity solution in eqs. (\ref{newNATD})-(\ref{newNATDrr}), with the directions along which the branes are suspended.}
\label{table:DoubleNATDBraneSetup}
\end{table}

We summarise the brane content of our solution in Table \ref{table:DoubleNATDBraneSetup}. From this table we see that all D-branes in our solution intersect each other only along the $t$ and $\rho$-directions. We therefore expect that the resulting worldvolume theory of the intersecting branes is a (1+1)-dimensional quantum field theory. However, since the D-branes have finite length along the $\rho$-direction, at low enough energies 
 this theory is described by a (0+1)-dimensional theory, a matrix model.

In conclusion, we have  two arrays of NS5-branes forming a grid in the $(\rho, \rho_1)$-plane. The first set, denoted by NS5, is located at $\rho_1 = \pi, 2\pi, \ldots, n_1 \pi$ and is extended along $(t,\alpha,\beta,\rho,\chi,\xi)$ whilst the second set, denoted by NS5', located at $\rho = \pi, 2\pi, \ldots, n \pi$ is extended along $(t,r,\beta,\rho_1,\chi_1,\xi_1)$. In addition, as a consequence of large gauge transformations, the D1 and D3'-branes expand onto fuzzy two-spheres with equilibrium radius of $\rho_1=n_1\pi$ giving rise to D3 and D5-branes. Moreover at any value of $\rho_1=n_1\pi$ we have usual D1 and D3' charge. We will see in the next section that these branes are indeed BPS.

To close this section, some comments about IIB matrix models are now in order.
Matrix models constructed upon the plane wave background in IIB supergravity have been studied in the literature following different approaches \cite{Ishibashi:1996xs,SheikhJabbari:2004ik, Lozano:2006jr, Sadri:2003mx, Bonelli:2002mb}. 
For instance, the author in  \cite{SheikhJabbari:2004ik} considered the quantisation of the 3-brane action whereas the starting point in \cite{Lozano:2006jr} was to consider the action for coincident gravitons \cite{Janssen:2002cf} for the maximally supersymmetric IIB pp-wave background.
In the former case, the model contains eight transverse non-Abelian scalars which explicitly realise the $SO(4)\times SO(4)$ symmetry of the pp-wave background whilst in the later
the fuzzy 3-sphere vacua were constructed as $S^1$ fibrations over fuzzy 2-spheres which break the isometries of the model down to  $(SO(3)\times U(1))^2$. 
The main difference of these models with the matrix model studied here is that in the present case there is only one set of three non-Abelian scalars which correspond to the (fuzzy) $\Omega_{2}(\chi_1,\xi_1)$
where the D1 and D3 expand into D3' and D5-branes respectively.  It would be interesting to understand if the matrix model studied here is embedded as a particular case of the above described matrix models.

\section{Dielectric branes}\label{sec:DielectricBranes}
In this section we will study D3 and D5 probe branes in the background of the NATD solution of eqs. (\ref{newNATD})-(\ref{newNATDrr}). We will show that such probe branes correspond to spherical branes carrying D1 and D3' charge respectively, and have radii $n_1 \pi$, proportional to their charge. This supports  the explanation of the brane set-up characterising the NATD solution studied in the preceding section.

\subsection{Dielectric D3 brane}\label{secd3}
Consider a D3-brane with worldvolume along $(t, \rho,\chi_1,\xi_1)$. Its induced metric is of the form
\begin{equation}
    ds^2_{D3} = L^2 \Big( -\cosh^2 r\;dt^2 + \frac{4 \alpha'^2}{L^4 \cos^2 \alpha}d\rho^2 +  \frac{4\alpha'^2\rho_1^2\sinh^2r}{\tilde{\Delta}}d\Omega_2^2(\chi_1, \xi_1) \Big).
\end{equation}
Large gauge transformations induce D1 charge proportional to $n_1\pi$ to this D3-brane, as can be seen from the WZ action for the D3-brane,
\begin{align}
	S^{WZ}_{D3} =& T_{D_3} \int_{\left(t, \rho, S^2_{\left( \chi_1,\xi_1 \right)} \right)} C_4 - C_2 \wedge B_2 =T_{D_3}\alpha' n_1\pi \int_{\left( t, \rho, S^2_{(\chi_1,\xi_1)} \right) }C_{2(t,\rho)}dt\, d\rho\, d\Omega_{2}(\chi_1,\xi_1), \nonumber\\
	=&T_{D_3}\frac{L^4 n_1\pi^2\sinh^4 r}{2} \int_{\mathbb{R}_t}  \int_{n\pi}^{(n+1)\pi} d\rho\;\rho ,
\end{align}
where $C_{2(t,\rho)}$ is the $(t,\rho)$ component of the $C_2$ potential in eq. (\ref{c2}). For the DBI action we find
\begin{align}
    S^{DBI}_{D3} &= T_{D_3} \int_{t, \rho, S^2_{_{(\chi_1,\xi_1)}}} d^{4}x\;e^{-\Phi} \sqrt{- \text{det} (g-\mathcal{F})}, \\
    &=T_{D_3} \frac{L^2\pi \sinh ^3r \cosh r}{8\alpha'}\int_{\mathbb{R}_t} \int_{n\pi}^{(n+1)\pi} d\rho\; \sqrt{16\alpha'^2 \rho^2 + L^4 \cos^4 \alpha} \\
    &\quad\times \sqrt{16\alpha'^2(\rho_1 - n_1\pi)^2\rho_1^2 + (n_1 \pi)^2 L^4 \sinh^4 r}. \nonumber
\end{align}
We easily see that the D3-brane finds its equilibrium position by sitting at the points $\alpha = \pi/2$, $\rho_1 = n_1 \pi$ where the DBI action reduces to 
\begin{align}
   S_{D3}^{DBI} = T_{D_3}\frac{L^4 n_1\pi^2\sinh r \cosh r }{2}\int_{\mathbb{R}_t} \int_{n\pi}^{(n+1)\pi} d\rho\;\rho.
\end{align}
We find that close to $r\sim\infty $ the D3-brane becomes BPS as the leading order terms in the DBI and WZ actions exactly cancel.
 In this limit the D3 extending on $S^2_{(\chi_1, \xi_1)}$ sits at the singularity at $\alpha=\pi/2$ and has a radius 
\begin{equation}\label{radius}
    R \sim \frac{2 n_1 \pi}{L \sinh r}.
\end{equation}
We can repeat the above computation for a probe D3'-brane along $(t, \rho,\chi,\xi)$. However, this does not correspond to a dielectric brane as the minimum of the potential felt by this D3'-brane sets its equilibrium radius, $\rho_1$, to zero. This is to be expected since the radius of the $S^2_{(\chi,\xi)}$ is $\rho$, which is part of the worldvolume coordinates of the D3'-brane. We will see in the next subsection however that D3'-branes can polarise into spherical D5-branes.

\begin{figure}[h!]
  \centering
  \includegraphics[width=11cm]{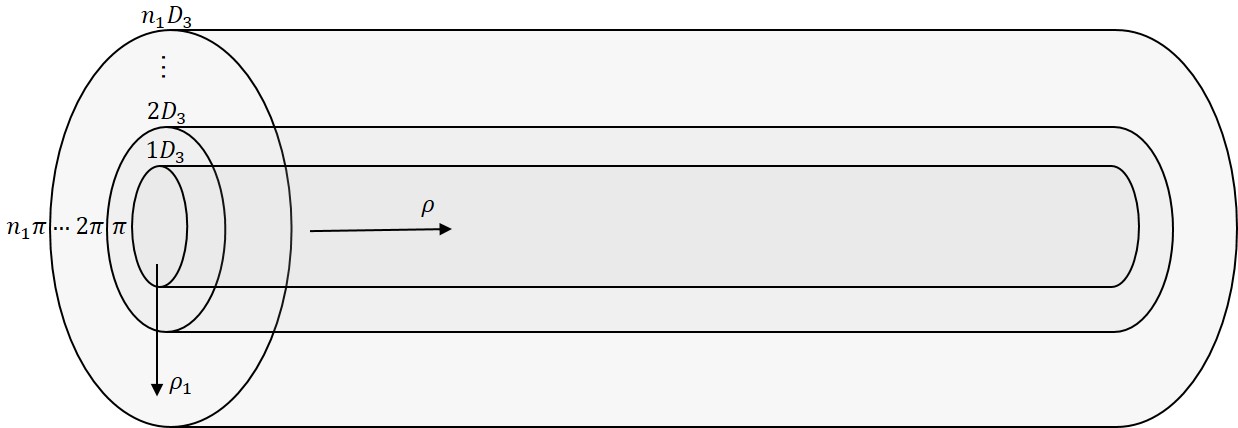}
  \caption{Expanded D3-branes on 2-spheres with radii $\rho_1 = n_1 \pi$.}
  \label{fig:ExpandedBranes}
\end{figure}

\subsection{Dielectric D5 brane}
Next we consider a probe D5-brane with worldvolume coordinates along $(t, \rho, \chi, \xi, \chi_1, \xi_1)$, wrapping the $S^2$ along ${(\chi_1, \xi_1)}$. The induced metric on this D5-brane is 
\begin{equation}
    ds^2_{D5} = L^2 \Big( -\cosh^2 r dt^2 + \frac{4\alpha'^2}{L^4 \cos^2 \alpha} d\rho^2 + \frac{4\alpha'^2\rho^2\cos^2\alpha}{\Delta}d\Omega_2^2(\chi, \xi) +  \frac{4\alpha'^2\rho_1^2\sinh^2r}{\tilde{\Delta}}d\Omega_2^2(\chi_1, \xi_1) \Big) 
\end{equation}
The DBI and WZ actions for this probe brane,  in the presence of large gauge transformations, are respectively 
\begin{align}
    S^{DBI}_{D5}
    &= -\frac{T_{D_5} L^2\pi^2}{2} \int_{\mathbb{R}_t} \int_{n\pi}^{(n+1)\pi}dt\;d\rho\;\cosh r \sinh r\;\sqrt{16\alpha'^2(\rho_1 - n_1\pi)^2\rho_1^2 + L^4 (n_1 \pi)^2 \sinh^4 r}\nonumber \\
    &\qquad \qquad \qquad   \times \sqrt{16\alpha'^2(\rho - n\pi)^2\rho^2 + L^4 (n \pi)^2 \cos^4 \alpha},\\
    S^{WZ}_{D5} &= T_{D_5}\int_{\left( t,\rho,S^2_{(\chi,\xi)},S^2_{(\chi_1,\xi_1)} \right)} \left( C_4 - C_2\wedge B_2+\frac{1}{2} C_2 \wedge \Delta B_2\right)\wedge \Delta B_2, 
\end{align}
where $\Delta B_2=n_1\pi d\Omega_{2}(\chi_1,\xi_1)+n\pi d\Omega_{2}(\chi,\xi)$. From the form of the WZ action we see that the D5-brane carries both D1 and D3'-brane charges. Their separate
contributions using the RR potentials (\ref{c2})-(\ref{c6}) are 
\begin{align}
 T_{D_5}\int \left( C_4 - C_2\wedge B_2\right)\wedge \Delta B_2&=- 2 (n_1 \pi) T_{D_5}L^4\pi^2\alpha'\sinh ^4r\int_{\mathbb{R}_t} \int_{n\pi}^{(n+1)\pi} d\rho\;\rho^2,\\
  T_{D_5}\int C_2\wedge \Delta B_2\wedge \Delta B_2&=2(n_1 \pi) T_{D_5}L^4 \pi^2 \alpha'\sinh ^4r\int_{\mathbb{R}_t} \int_{n\pi}^{(n+1)\pi} n\pi\;d\rho\;\rho.
\end{align}
We see that the charge of both terms is proportional to $n_1\pi$. The D5-brane is in equilibrium at $\rho_1= n_1 \pi, \alpha=\frac{\pi}{2}$ and $r \sim \infty$ when both the WZ and DBI actions are of the same form and cancel each other. 
However, notice that at $\alpha=\pi/2$ the $\Omega_2(\chi,\xi)$ shrinks to zero size and the D5-brane becomes effectively the D3-brane studied in section \ref{secd3}. Moreover, if we instead take $n=0$ together with $\rho_1= n_1 \p$ and $r \sim \infty$ the D5-brane is indeed stable because of its D3' charge. The D5-brane therefore wraps an $S^2_{(\chi_1, \xi_1)}$ of radius (\ref{radius}) carrying D3' charge proportional to its radius.

\section{Double Abelian (Hopf)T-Dual of the $AdS_5\times S^5$ solution }\label{DATD}
In this section we study the solution that is obtained after the application of two Abelian T-dualities (ATD) along the Hopf-fibre angles inside the $S^3$ in both the $AdS_5$ and $S^5$ subspaces of the solution in eqs. (\ref{adsspace}) and (\ref{flux}). We will demonstrate that when an ATD is only applied to the first of these two subspaces, the resulting background belongs to the  Lin-Maldacena classification of half-BPS geometries in type IIA supergravity \cite{Lin:2005nh}. We will also identify 
the solution after two ATD's as a double scaling limit of the NATD solution discussed in Section \ref {sec:DoubleNATDBackground}.

\subsection{ATD along the $S^3\subset AdS_5$}\label{subsec:ATDonAdS5}
Let us consider an ATD along the $\psi_1$ coordinate of $S^{1}_{\psi_1}\subset S^3\subset AdS_5$. Following the usual Buscher rules, see for instance \cite{Hassan:1999bv},  we find, 
by replacing $(\theta_1,\phi_1)\rightarrow (\chi_1,\xi_1)$, the type IIA background
\begin{align}\label{first}
ds^2&=L^2\left(-\cosh^2r dt^2+dr^2+\frac{4 \alpha'^2}{L^4\sinh^2r}d\tilde{\psi}_1^2+\frac{\sinh^2r}{4}d\Omega_{2}^2(\chi_1,\xi_1)+d\Omega_5^2 \right),\\
&\qquad B_2=\alpha'\tilde{\psi}_1 \,d\Omega(\chi_1,\xi_1)+d\Lambda, \quad \Lambda=\alpha'\tilde{\psi}_1\cos \chi_1 d\xi_1 ,\qquad e^{-2\Phi}=\frac{L^2\sinh^2r}{4\alpha'},\\
&\qquad\qquad \qquad F_4= - \frac{L^4}{2\sqrt{\alpha'}}\cosh r\sinh^3 r\;dr\wedge dt \wedge d\Omega(\chi_1, \xi_1)\label{last}. 
\end{align}
where the T dual coordinate $\tilde{\psi}_1$, is now periodic with period $\pi$  \cite{Lozano:2016kum}. 
Since T-duality acted on the $AdS_5$ symmetries
we have broken the $SO(4,2)$ conformal symmetry down to $\mathbb{R}_t\times SO(3)$. This solution breaks half of the supersymmetries. The background in eq. (\ref{first}) is singular at $r=0$. As in the previous cases, this is due to the presence of NS5-branes, as can be seen by computing the flux of $H_3$ through $(\tilde{\psi}_1,S^2_{(\chi_1,\xi_1)})$,
\begin{equation}\label{fluxh}
N_{_{NS5}}=\frac{1}{4\pi^2\alpha'}\int_{\tilde{\psi}_1,S^2_{(\chi_1,\xi_1)}}H_3=1. 
\end{equation}
where the period of the T-dual coordinate is proportional to the number of NS5-branes. Due to the non-trivial topology of the $(\chi_1, \xi_1)$ cycle at $r=0$ we allow the presence of large gauge transformations which wind $n_1$-times around the one-cycle defined by $\tilde{\psi}_1$, analogous to section \ref{subsec:QuantisedCharges}. If we allow $\tilde{\psi}_1$ to lie in $[0,n_1\pi]$ we find 
$N_{_{NS5}}=n_1$. In addition we also have $N_{D_2}$ units of RR flux
\begin{equation}\label{nd2}
N_{D_2}=\frac{1}{(2\pi\sqrt{\alpha'})^5}\int_{\tilde{\psi}_1,\Omega_5} \star F_4=\frac{L^4}{8\pi \alpha'^2}.
\end{equation}
The above picture suggests that we have a periodic arrangement of NS5-branes, equally separated along the $\tilde{\psi}_1$ direction with $N_{D_2}$ units of flux through $(\psi_1,\Omega_5)$ at each interval comprised by two consecutive parallel NS5-branes.
 In the next subsection we shall prove that this solution (\ref{first})-(\ref{last}) is dual to $\mathcal{N}=4$ SYM on $\mathbb{R}_{t}\times S^3/\mathbb{Z}_{n_1}$ and can therefore be understood in terms of the classification of type IIA geometries studied in \cite{Lin:2005nh}.

.
\subsubsection{The ATD as a Lin-Maldacena background}\label{subsec:ATDLM}
In \cite{Lin:2005nh} Lin and Maldacena found the gravity duals of half-BPS states of the $\mathcal{N}=4$ SYM theory on $\mathbb{R}_t\times S^3$, the dual field theories of which are obtained as truncations of the modes
on $S^3$ by either $SU(2)_L\subset SO(4,2)$ or subgroups of it.
The type IIA geometries with $\mathbb{R}_t\times SO(3)\times SO(6)$ isometries correspond to particular cases of a more general class of half-BPS geometries in IIB supergravity and M-theory, discussed in \cite{Lin:2004nb}, and are fully characterised by a two dimensional potential function $V(\sigma,\eta)$ solving a Laplace equation. The problem of finding type IIA solutions belonging to this class then boils down to solving an electrostatic problem for this potential function subject to certain boundary conditions. These generic IIA gravity duals are explicitly given by
 \begin{equation}\label{generic}
 \begin{split}
ds_{10}^2=&\left(\frac{\ddot V-2\dot V}{-V''}\right)^{1/2}\left[-4\frac{\ddot V}{\ddot  V-2\dot V}dt^2+2\frac{V'' \dot V}{\Delta}d\Omega_2^2+\frac{-2V''}{\dot V}(d\sigma^2+d\eta^2)+4d\Omega_5^2\right],\\
&\qquad\qquad e^{4 \phi} = \frac{4(2 \dot V - \ddot V)^3}{V'' \dot V^2  \Delta^2}, \qquad B= 2 \left(\frac{\dot V \dot V'}{ \Delta} + \eta \right) d \Omega_2,\\
&\qquad C_{1} = \frac{2\dot V \dot V'}{2 \dot V - \ddot V} dt, \qquad C_{3} = - \frac{4 \dot V^2 V''}{ \Delta} dt \wedge d\Omega_2, \quad F_4=dC_3,\\
&\qquad\qquad\qquad\qquad\quad \Delta=(\ddot V-2\dot V)V''-\dot{V}'^{2},
\end{split}
\end{equation}
where $\dot{V}=\sigma\partial_{\sigma} V$, $V'=\partial_{\eta}V$ and $V(\sigma,\eta)$ solves the equation
\begin{equation}
\ddot{V}+\sigma^2 V''=0.
\end{equation}
In order to find regular solutions the potential function $V(\sigma,\eta)$ is subject to boundary conditions determined by a configuration of parallel conducting disks along the $\eta$-axis at positions $\eta=\eta_{i}$. These conditions were ascertained by requiring that either the $S^2$ or the $S^5$ cycles shrink smoothly to zero size. The quantisation of fluxes through non-trivial six and three-cycles in the geometry determines a relation between D-brane charges and the parameters in the electrostatic configuration. In concrete, the distance between two consecutive disks is associated with the NS5-brane charge via $\eta_i=\pi/2 N_5^{(i)}$ whilst their charge is related to the number of D2-branes as $Q_{i}=\pi^2/8 N_2^{(i)}$.
\cite{Lin:2005nh}.  Moreover, the well-definiteness of the metric components in the metric of eq. (\ref{generic}) imposes that the conducting disks are immersed in a background potential $V_{b}$ characterising the asymptotic behaviour of the solution. The general form of the solution is therefore $V=V+V_{b}$.

Let us consider the particular case of $\mathcal{N}=4$ $U(N)$ SYM on $\mathbb{R}_{t}\times S^3/\mathbb{Z}_{n_{1}}$. 
On the $S^3$, all modes are massive except the holonomy of the gauge field around the non-trivial cycle of $S^3/\mathbb{Z}_{n_{1}}$. 
We use gauge symmetry and the condition $U^{n_1}=1$ to  write the holonomy as $U=\textrm{diag}(1,\omega^{1},\ldots,\omega^{1},\omega^{2},\ldots,\omega^{I})$. 
where  $\omega=e^{i\frac{2\pi}{n_1}}$ and $I=0,\ldots n_1-1$.
All in all, a vacuum of the theory is characterised by $n_1$ integers and the multiplicities $\tilde{N}_{I}$ specifying the number of times the $\omega^{l}$ entry appears in the diagonal, such that
\begin{equation}
N=\sum_{I=0}^{n_1}\tilde{N}_I.
\end{equation}
In particular,  the $\mathbb{Z}_{n_1}$  invariant vacuum is defined by setting $\tilde{N}_I=\frac{N}{n_{1}}$  for all $I$. 
On the supergravity side, the solution dual to $\mathbb{R}_{t}\times S^3/\mathbb{Z}_{n_{1}}$ is characterised by the potential function \cite{Lin:2005nh}
\begin{equation}\label{pot}
V=-2\log\sigma+V_{b},\qquad V_{b}=\sigma^2-2\eta^2.
\end{equation}
The electrostatic picture of this solution corresponds to a periodic arrangement of disks along the $\eta$ axis \cite{Lin:2005nh}. 

In order to see that the solution in  eqs. (\ref{first})-(\ref{last}) corresponds to the above described configuration
consider the following change of coordinates.
\begin{equation}\label{transf}
 \sigma=\cosh r,\quad \eta=\frac{2\alpha' }{L^2}\tilde{\psi}_1.
 \end{equation}
 
 The solution then reads 
\begin{equation}
\begin{split}
ds^2=&\frac{L^2}{4}\left(-4\sigma^2 dt^2+\frac{4}{\sigma^2-1}(d\sigma^2+d\eta^2) +(\sigma^2-1)d\Omega_2^2+4d\Omega_5^2\right),\\
&\quad B_2=\frac{L^2}{4}2\eta \,d\Omega_2(\chi_1,\xi_1),\qquad e^{-2\Phi}=\frac{L^2}{4\alpha'}(\sigma^2-1),\\
&\qquad F_4=-\frac{L^{4}}{16\alpha'}8\sigma(\sigma^2-1)d\sigma\wedge dt\wedge d\Omega_{2}(\chi_1,\xi_1), 
\end{split}
\end{equation}
which is the solution we would have obtained by plugging the potential function\footnote{up to some rescalings} in eq. (\ref{pot}) in the generic IIA solution in eq. (\ref{generic}).

To close this subsection, notice that the solution in eqs.  (\ref{first})-(\ref{last}) can be analytically continued using $r=i\alpha, t\rightarrow \phi$ and $L\rightarrow i L$. The resulting theory is the T dual of $AdS_5\times S^5/Z_{k}$ 
the field theory dual of which corresponds to an $\mathcal{N}=2$ supersymmetric quiver field theory \cite{Fayyazuddin:1999zu, Witten:1998qj}. This relation was anticipated in \cite{Lin:2004nb} and is in the same spirit as the relation between the solutions discussed in sections \ref{sfetsos-thompson} and \ref{LNsolution}
found in \cite{Lozano:2017ole}. The T-dual of $AdS_5\times S^5/Z_{k}$  was studied in \cite{Lozano:2016kum} in the context of the Maldacena-Gaiotto class of geometries.

\subsection{Double Abelian T-duality}\label{subsec:DATD}
Let us now consider the effect of a second ATD along the $\psi$ coordinate inside $S^5$ in the solution in eq. (\ref{first})-(\ref{last}). Following the rules in \cite{Hassan:1999bv} we find a solution of type IIB supergravity which, by replacing $(\theta,\phi)\rightarrow (\chi,\xi)$, reads 
\begin{align}\label{double}
ds^2&=L^2\left(-\cosh^2r\;dt^2+dr^2+\frac{4 \alpha'^2}{L^4\sinh^2r}d\tilde{\psi}_1^2+\frac{\sinh^2r}{4}d\Omega_{2}^2(\chi_1,\xi_1)\right.\nonumber\\
&\left.d\alpha^2+\sin^2\alpha\;d\beta^2+\frac{4\alpha'^2}{L^4\cos^2\alpha}d\tilde{\psi}^2+\frac{\cos^2\alpha}{4} d\Omega_2^2(\chi,\xi) \right)\\
B_2&=\alpha'\tilde{\psi} \,d\Omega(\chi,\xi)+\alpha'\tilde{\psi}_1 \,d\Omega(\chi_1,\xi_1),\qquad e^{-2\Phi}=\frac{L^4\cos^2\alpha\sinh^2r}{(4\alpha')^2},\\
&\quad F_5=\frac{L^4}{2}\left(\cosh r\sinh^3 r\;dt\wedge dr \wedge d\tilde{\psi} \wedge d\Omega_2(\chi_1, \xi_1) \right.\nonumber\\
&\qquad \quad\left. +\sin\alpha \cos^3\alpha\;d\alpha\wedge d\beta\wedge d\tilde{\psi}_1 \wedge d\Omega_2(\chi, \xi) \right)
\end{align}
where $\tilde{\psi} \in [0,\pi]$. From the analysis in Appendix \ref{sec:susy}, this solution preserves 8 supercharges. There are two singular points in the solution at $r=0$ and $\alpha=\frac{\pi}{2}$ due to the presence of two sets of NS5-branes, as can be seen from the flux of $H_3$ through the two three-cycles $\Sigma'_3=(\xi_1,\chi_1,\tilde{\psi}_1)\vert_{r=0}$ and $\Sigma_3 =(\xi,\chi,\tilde{\psi})\vert_{\alpha=\pi/2}$,
\begin{equation}
N_{_{NS5_{1}}}=\frac{1}{4\pi^2 \alpha'}\int_{\Sigma'_3}H_3=1, \qquad N_{_{NS5_{2}}}=\frac{1}{4\pi^2 \alpha'}\int_{\Sigma_3}H_3=1. 
\end{equation}
In addition, the D3 Page charge for this double ATD solution is
\begin{equation}\label{chargeATD}
    N_{D_3}^{ATD}=\frac{1}{(2\pi \sqrt{\alpha')^2}}\int F_5,\quad \quad  L^4=16\pi \alpha'^2 N_{D_3}^{ATD}.
\end{equation}
As in section \ref{subsec:QuantisedCharges}, taking into account large gauge transformations in the non-trivial cycles $\Sigma_3$ and $\Sigma'_{3}$ we find 
$N_{_{NS5_{1}}}=n_1$ and $N_{_{NS5_{2}}}=n$.
The brane set-up characterising this solution consists of two periodic configuration of NS5-branes, $NS5_{1}$ and $NS5_2$, at positions  $\pi, 2\pi \ldots$, along  the $\tilde{\psi}_1, \tilde{\psi}$ circles respectively with $n_1N_{D_3}$ D3-branes stretched along the $\psi$-direction between the NS$5_{1}$-branes.

Let us now compare this solution with the one studied  in section \ref{sec:DoubleNATDBackground}, in the double limit $\rho,\rho_1\rightarrow \infty$. It is clear that in this limit the leading order behaviour of the metric and the $B_2$ field of the solutions in eqs. (\ref{newNATD}) and (\ref{double}) exactly agree if we identify $\rho,\rho_1\rightarrow \tilde{\psi},\tilde{\psi}_1$. 
To be more precise, in order to match both solutions globally we should consider the theory on the interval $\rho\in [n\pi,(n+1)\pi]$ and $\rho_1\in [n_1\pi,(n_1+1)\pi]$ and send $n,n_1\rightarrow\infty$ \cite{Lozano:2016kum}. 
There is no matching for the RR fields and the dilaton. More explicitly we find the relations
\begin{equation}\label{relationsa}
\lim_{\substack{\rho\rightarrow \infty \\ \rho_1\rightarrow \infty}}\{ds^{2}_{_{\textrm{NATD}}},B_{_{\textrm{NATD}}},e^{\Phi_{_{\textrm{NATD}}}}, (e^{\Phi}F_{3,5})_{_{NATD}}\}=\{ds^{2}_{_{\textrm{ATD}}},B_{_{\textrm{ATD}}},\rho \rho_1e^{\Phi_{_{\textrm{ATD}}}},(e^{\Phi}F_{3,5})_{_{ATD}}\}.
\end{equation}
Notice that, compared to the relations earlier found between ATD and NATD solutions, see e.g. \cite{Lozano:2016kum, Lozano:2016wrs}, an extra $\rho_1$ factor appears in the dilaton due to the second NATD.
We will now demostrate that the relation between NATD and ATD solutions 
can be made precise by considering a double scaling limit in the NATD solutions. 
To see this explicitly, we first change coordinates according to $\rho=\rho_0+\psi$ and consider the limits 
\begin{equation}\label{finalscaling}
 \rho_0\rightarrow \infty, \quad g_{s}^{\textrm{\tiny NATD}}\rightarrow \infty,\quad \frac{\rho_0}{g_{s}^{\textrm{\tiny NATD}}}=\textrm{fixed}\equiv \frac{1}{g_{s}^{\textrm{\tiny ATD}}}.
\end{equation}
By recovering the appropriate factors of $g_s$ in the solutions of eqs. (\ref{newNATDrr})-(\ref{newNATD}) and (\ref{double}), we have verify that the  the double scaling limit in eq. (\ref{finalscaling}) leads to 
an exact agreement between NATD and ATD solutions\footnote{Notice that for the double NATD solution we have to consider in addition a corresponding limit for $\rho_1$.}. 
This suggest, though we did not verified it explicitly, that this double asymptotic limit will relate more general SU(2) NATD and ATD solutions studied in the literature.

\section{Summary and Conclusions}\label{sec:Conclusion}
In this work we constructed new solutions in type IIB supergravity that are obtained by applying (non)-Abelian T-dualities on the maximally supersymmetric $AdS_5\times S^5$ solution of type IIB supergravity. We started in section \ref{sec:AdssxS5andNATDSolutions} by briefly reviewing the solutions that are obtained by applying a single NATD along the SU(2) isometries inside either the $AdS_5$ or $S^5$ subspaces. Acting with a second NATD on the residual SU(2) isometries of the above solutions in section \ref{sec:DoubleNATDBackground}, we obtained an $\mathbb{R}_t \times SO(3)\times SU(2)\times U(1)$ symmetric solution in type IIB preserving eight supercharges. The study of the Page charges in section \ref{subsec:QuantisedCharges} showed that 
our solution is characterised by two sets of NS5-branes, denoted by NS5 and NS5',  that form a grid in the $(\rho,\rho_1)$-plane, see Table \ref{table:DoubleNATDBraneSetup}. In addition there are D1-branes that extend along the $\rho$-direction between the NS5-branes and, due to the presence of large gauge transformations of the $B_2$ field, are blown up into D3-branes which wrap around the $\Omega_2(\chi_1, \xi_1)$-sphere. Besides this there is a second set of D3-branes along $\rho$ and $\Omega_2(\chi, \xi)$ that due to the same mechanism blow up into D5-branes which also wrap around $\Omega_2(\chi_1, \xi_1)$. 
Since all the D-branes intersect only in the $(t,\rho)$ directions and have finite length along $\rho$, the low energy limit of our brane set-up is described by a matrix model characterised by fuzzy sphere vacua. 
By studying probe branes in section \ref{sec:DielectricBranes} we showed that D3 and D5-branes have polarised from D1 and D3' branes respectively with radii proportional to their charge.
As for other NATD solutions studied in the literature, it would be interesting to understand if this solution corresponds to a more general class of IIB supergravity, matrix models, preserving $\mathbb{R}_t \times SO(3)\times SU(2)\times U(1)$ isometries
and eight supercharges. 

Moreover, we showed in sections \ref{subsec:ATDonAdS5}-\ref{subsec:ATDLM} that the solution generated after ATD along the Hopft-fibre angle on the $S^3$ inside $AdS_5$ belongs to the half-BPS class of geometries studied by Lin and Maldacena \cite{Lin:2004nb}
and is dual to $\mathcal{N}=4$ theory on $\mathbb{R}\times S^3/\mathbb{Z}_k$. A subsequent application of an ATD on the Hopft-fibre angle on the $S^3$ inside $S^5$ bring us back to type IIB supergravity. This solution preserves eight supercharges and is characterised by two periodic arrangements of NS5-branes, NS$5_1$ and NS$5_2$, along the dualised coordinates with $n_1N_{D_3}$ D3-branes stretched
between consecutive NS$5_1$-branes. In addition, we showed that this solution is related to the (double) NATD solution studied in Section \ref{sec:DoubleNATDBackground} via the limit where $\rho,\rho_1\to\infty$. 
We improved this argument by showing that via a double scaling limit  the relation between NATD and ATD solutions is exact for the solutions studied in this work.
The relation between NATD and ATD solutions was also studied in the particular case of the NATD and ATD solutions studied in sections \ref{LNsolution} and \ref{subsec:ATDonAdS5}
 by taking a limit in the potential function, in the Lin-Maldacena language \cite{Lin:2005nh}, characterising these half-BPS solutions. We found that
 the relation via the double scaling limit of these solutions is a consequence of the result found in \cite{Ishiki:2006yr} where it was shown that any vacuum of the theory on $\mathbb{R}\times S^3/\mathbb{Z}_k$ is embedded in a particular vacua of the PWMM.

\acknowledgments
We would like to thank Carlos N\'{u}{\~n}ez and Yolanda Lozano for many useful discussions throughout the completion of this work. The work of J.v.G. is supported by an STFC scholarship, S. Z. is a Newton International Fellow of the Royal Society.
\appendix

\section{Comments about the supersymmetry of the solution}\label{sec:susy}
In this appendix we will demonstrate that the  $AdS_5\times S^5$ Killing spinor breaks supersymmetry when it is constrained to be independent 
of the $SU(2)$ angles inside the $AdS_5$ and $S^5$ subspaces. 
 A procedure along these lines was already performed in \cite{Macpherson:2014eza} for the case where the spinor is independent of the $SU(2)$ angles inside $S^5$.
 Here we will generalise that argument by imposing also the independence on the $SU(2)$ angles inside $AdS_5$.
 Following the results of \cite{Kelekci:2014ima}, 
 the number of supersymmetries preserved by this spinor will correspond to the ones after applying two NATD along the $SU(2)$ isometries of the $AdS_5\times S^5$ solution.

We first introduce a convenient set of frame fields for the solution in eq. (\ref{adsspace}) which explicitly exhibit the SU(2) isometries of the background. They are 
\begin{equation}
\begin{split}
e^{t}=&L\cosh r\;dt,\quad e^{r}=L\;dr,\quad e^{a}=\frac{1}{2}L\sinh r\,  \omega_{a},\qquad  a=3,4,5,\\
e^{\alpha}=&L\; d\alpha,\quad e^{\beta}=L \sin \alpha\;d\beta,\quad e^{i}=\frac{1}{2}L\cos\alpha\, \sigma_{i},\qquad  i=6,7,8.
\end{split}
\end{equation}
where $\omega_{i}$ and $\sigma_{i}$ were defined in eq. (\ref{mcforms}). 
In this basis the RR five form in eq. (\ref{flux}) reads 
\begin{equation}
F_{5}=\frac{4}{L}\left(e^{t,r,\theta_1,\phi_1,\psi_1}-e^{\theta,\phi,\psi,\alpha,\beta}\right).
\end{equation}
Supersymmetric solutions impose the conditions\footnote{were we have written the spinor in complex notation as  $\epsilon=\epsilon_1+i\epsilon_2$} 
 \begin{align}
& \left(\nabla_{\mu}+\frac{i}{2 L}\Gamma^{tr\theta_1 \phi_1 \psi_1}\Gamma_{\mu}\right)\epsilon=0, \qquad \mu=t,r,\theta_1, \phi_1, \psi_1,\label{susy11}\\
  &\left(\nabla_{\nu}-\frac{i}{2 L}\Gamma^{\theta \phi \psi \alpha \beta}\Gamma_{\nu}\right)\epsilon=0, \qquad \quad \nu=\theta, \phi, \psi, \alpha, \beta,\label{susy22}
 \end{align}
 where the Majorana-Weyl spinor $\epsilon$ satisfies
\begin{equation}
\Gamma^{tr\theta_1 \phi_1 \psi_1 \theta \phi \psi \alpha \beta }\epsilon=-\epsilon.
\end{equation}
 We are interested in solutions to eqs. (\ref{susy11}) and (\ref{susy22}) which are consistent with the independence of the spinor on the $SU(2)$ angles in both $AdS_5$ and $S^5$.
 By imposing these conditions we find the equations 
    \begin{equation}
  \begin{split}
  &(\Gamma^{\phi_1\psi_1}-\cosh r \Gamma^{r,\theta_1}-i \sinh r \Gamma^{tr\phi_1\psi_1})\epsilon=0, \quad (\Gamma^{\phi\psi}-\sin\alpha \Gamma^{\theta,\psi}-i \cos\alpha \Gamma^{\phi\psi\alpha\beta})\epsilon=0,\\
    &(\Gamma^{\theta_1\psi_1}+\cosh r \Gamma^{r,\phi_1}-i \sinh r \Gamma^{tr\theta_1\psi_1})\epsilon=0,\quad (\Gamma^{\theta\psi}+\sin\alpha \Gamma^{\phi,\alpha}-i \cos\alpha \Gamma^{\theta\psi\alpha\beta})\epsilon=0,\\
        &(\Gamma^{\theta_1\phi_1}-\cosh r \Gamma^{r,\psi_1}-i \sinh r \Gamma^{tr\theta_1\phi_1})\epsilon=0,\quad (\Gamma^{\theta\phi}-\sin\alpha \Gamma^{\psi,\alpha}-i \cos\alpha \Gamma^{\theta\phi\alpha\beta})\epsilon=0,\\
         &(2\partial_{t}+\sinh r \Gamma^{tr}-i\cosh r\Gamma^{r\theta_1\phi_1\psi_1})\epsilon=0, \qquad (2\partial_{\beta}-\cos\alpha\Gamma^{\alpha\beta}-i\sin\alpha \Gamma^{\theta\phi\alpha\beta})\epsilon=0,\\
  & (2\partial_{r}+i\Gamma^{t\theta_1 \phi_1 \psi_1})\epsilon=0, \qquad \quad \qquad \quad \qquad \quad  \quad (2\partial_{\alpha}+i\Gamma^{t\theta\phi\psi\beta})\epsilon=0,
  \end{split}
 \end{equation}
 These equations can be massaged giving rise to the following set of equations 
 \begin{align}
&2\partial_{\beta}\epsilon+i\epsilon=0, \qquad\qquad\quad   2\partial_{t}\epsilon-i\epsilon=0,\\
&2\partial_{\alpha}\epsilon+i\Gamma^{\theta\phi\psi\beta}\epsilon=0, \qquad ~2\partial_{r}\epsilon-i\Gamma^{t\theta_1\phi_1\psi_1}\epsilon=0,
\end{align}
and the projector conditions 
\begin{align}
\Gamma^{\alpha\beta}\epsilon&=(-i\cos\alpha +\sin\alpha \Gamma^{\theta\phi\psi\beta})\epsilon=ie^{i\alpha \Gamma^{\theta\phi\psi\beta}}\epsilon ,\\
\Gamma^{r\theta_1\phi_1 \psi_1}\epsilon&=(-\cosh r+i\sinh r\Gamma^{t\theta_1\phi_1\psi_1})\epsilon=-e^{i t \Gamma^{\theta\phi\psi\beta}}\epsilon.
\end{align}
The Killing spinor is easily found and reads 
\begin{equation}
\epsilon =e^{\frac{i}{2}(t-\beta)}e^{\frac{i}{2}(r\Gamma^{t\theta_1 \phi_1 \psi_1}-\alpha \Gamma^{\theta \phi \psi \beta} )}\epsilon_{0},
\end{equation}
where $\epsilon_0$ is a spinor satisfying the projections 
\begin{equation}
\Gamma^{\alpha\beta}\epsilon_0=-i\epsilon_0,\qquad  \Gamma^{r\theta_1\phi_1 \psi_1}\epsilon_{0}=-\epsilon_{0}.
\end{equation}
Since these projectors commute with each other, we have a total of eight supercharges preserved by the solution. Had we started 
by imposing independence of the spinor on the Hopf-fibre angles of the $S^{3}$ inside $AdS_5$ and $S^5$, $\partial_{\psi_1}\epsilon=0$, $\partial_{\psi}\epsilon=0$, we would have obtained  
the same set of projections, ensuring that the solution after two Abelian T-dualities along the Hopf-fibre angles preserves also eight supercharges \cite{Hassan:1999bv}.

\section{A double scaling Limit for the NATD solutions}\label{scaling}
In this appendix we will consider a double scaling limit relating the solutions studied in sections \ref{LNsolution} and \ref{subsec:ATDonAdS5}. 
The relation between NATD and ATD solutions has been shown explicitly for the solutions studied in \cite{Lozano:2016wrs, Lozano:2016kum,Lozano:2017ole, Itsios:2017cew}.
However, as we showed in section \ref{subsec:DATD}, this relation is not exact since the dilaton and the RR fields do not match in this limit (see eq. (\ref{relationsa})).
Here we will show that the relation between NATD and ATD solutions is indeed exact by considering a double scaling limit. 
 We will prove this, in particular, for the solutions studied in sections \ref{LNsolution} and \ref{subsec:ATDonAdS5} by making contact 
 with the fact that both solutions belong to the Lin-Maldacena half-BPS class of geometries. To be more specific, 
 We will demonstrate that the potential functions characterising these two solutions are related via a double scaling limit, in analogy to the one considered in  \cite{Ling:2006up}.
 
Similarly to the analysis in section \ref{subsec:ATDLM}, it was shown in  \cite{Lozano:2017ole} that the NATD solution in eqs. (\ref{natd2}) is a half-BPS solution of the Lin-Maldacena class, the dual theory of which corresponds to a particular vacuum of the PWMM. In order to see this explicitly, consider the transformation of coordinates written in eq. (\ref{transf}) with the replacement $\psi_1\rightarrow \rho_1$. One can easily see that the corresponding supergravity solution is generated, using eq. (\ref{generic}),  by the potential function
\begin{equation}\label{bbmn}
V_{_{\textrm{NATD}}}=V_{0}(\sigma^2\eta-\frac{2}{3}\eta^3)-2V_0\eta\log\,\sigma=V_{_{\textrm{PWMM}}}-2V_0\eta\log\,\sigma.
\end{equation}
where $V_{_{PWMM}}$ is the background potential characterising the supergravity solutions dual to the PWMM
and $V_0$ is a constant. In the electrostatic picture, the configuration described by the potential function in eq. (\ref{bbmn}) corresponds to a {\it coarse grained} approximation of parallel conducting disks in the presence of an infinite conducting disk at $\eta=0$. 
The charge density of the continuous distribution is \cite{Lozano:2017ole} 
\begin{equation}\label{chargeq}
Q=\frac{L^4}{16\alpha'^2}V_{0}\eta.
\end{equation}
We now `zoom in' on a region close to $\eta_0$ by taking $\eta\rightarrow \eta_0+\hat{\eta}$.  The potential function in eq. (\ref{bbmn}) then becomes 
\begin{equation}
\hat{V}_{_{\textrm{NATD}}}=V_0\eta_0(\sigma^2-2\hat{\eta}^2)+V_0(\hat{\eta}\sigma^2-\frac{2}{3}\hat{\eta}^3)-2V_0\hat{\eta}\log\,\sigma-2V_0\eta_0\log\,\sigma,
\end{equation}
where we have ignored constant and linear terms in $\hat{\eta}$ that do not contribute to the supergravity solution. 
We then blow up the region close to $\eta_0$ by taking $\eta_0\rightarrow \infty$. We see from eq. (\ref{chargeq}) that in order to have a non-trivial configuration 
with finte charge we ought to take in addition $V_0\rightarrow 0$ such that $\hat{V_0}=V_0\eta_0=\textrm{fixed}$.
We get in this limit 
\begin{equation}
\hat{V}_{_{\textrm{NATD}}}\rightarrow \hat{V}_0((\sigma^2-2\eta^2)-2\log\,\sigma)= V_{_{\textrm{ATD}}},
\end{equation}
where $V_{_{\textrm{ATD}}}$ is the potential function for the solution we studied in section \ref{subsec:ATDLM}, see eq. (\ref{pot}).
Therefore, in the double scaling limit described above, the vacuum of the BMN matrix model studied in \cite{Lozano:2017ole} maps to any vacuum of the theory on $\mathbb{R}\times S^3/\mathbb{Z}_{k}$, the electrostatic picture of which corresponds to a periodic arrangement of equally spaced conducting disks along the $\hat{\eta}$ direction.
This is a supergravity realisation of the result found in \cite{Ishiki:2006yr}, previously anticipated in \cite{Lin:2005nh},  where it was shown that every vacuum of $\mathcal{N}=4$
SYM on $\mathbb{R}\times S^3/\mathbb{Z}_{k}$ corresponds to a particular vacuum of the PWMM.

In terms of the parameters in the NATD supergravity solutions, the double scaling limit corresponds to 
\begin{equation}
\rho=\rho_0+\psi,\quad \rho_0\rightarrow \infty, \quad g_{s}^{\textrm{\tiny NATD}}\rightarrow \infty,\quad \frac{\rho_0}{g_{s}^{\textrm{\tiny NATD}}}=\textrm{fixed}\equiv \frac{1}{g_{s}^{\textrm{\tiny ATD}}}.
\end{equation}


\end{document}